\documentclass[12pt]{article}
\usepackage{epsfig,amsfonts,amssymb,amsbsy,amsbsy,array}
\def\cvp{\raise 2pt\hbox{,}}

\def\tr{\mathop{\rm tr}\nolimits}

\def\d{{\rm d}}
\def\suN{{\rm SU}(N)}
\font\HUGE=cmr12 at 40pt
\def\plb#1#2#3{{\it Phys.\ Lett.\ }{\bf B #1} (#2) #3}
\def\npb#1#2#3{{\it Nucl.\ Phys.\ }{\bf B #1} (#2) #3}
\def\npps#1#2#3{{\it Nucl.\ Phys.\ Proc.\ Suppl.\ }{#1} (#2) #3}
\def\prl#1#2#3{{\it Phys.\ Rev.\ Lett.\ }{\bf #1} (#2) #3}
\def\jhep#1#2#3{{\it J. High Energy Phys.\ }{\bf #1} (#2) #3}
\def\prd#1#2#3{{\it Phys.\ Rev.\ }{\bf D #1} (#2) #3}
\def\atmp#1#2#3{{\it Adv.\ Theor.\ Math.\ Phys.\ }{\bf #1} (#2) #3}
\def\cmp#1#2#3{{\it Comm.\ Math.\ Phys.\ }{\bf #1} (#2) #3}
\def\pr#1#2#3{{\it Phys.\ Rep.\ }{\bf #1} (#2) #3}

\def\ijmpa#1#2#3{{\it Int.\ J.\ Mod.\ Phys.\ }{\bf A #1} (#2) #3}

\begin{document}
%
%\pagenumbering{roman}
%
\pagestyle{empty}
{\parskip 0in
\hfill PUPT-2027

\hfill NEIP-02-003

\hfill LPTENS-02/18

\hfill hep-th/0205171}

\vfill
\begin{center}
{\sffamily\large\bfseries FOUR DIMENSIONAL NON-CRITICAL STRINGS}
%\medskip

\vspace{0.4in}

Frank F{\scshape errari}{\renewcommand{\thefootnote}{$\!\!\dagger$}
\footnote{On leave of absence from Centre 
National de la Recherche Scientifique, Laboratoire de Physique 
Th\'eorique de l'\'Ecole Normale Sup\'erieure, Paris, France.}}\\
\medskip
{\it Joseph Henry Laboratories\\
Princeton University, Princeton, New Jersey 08544, USA\\ and\\
Institut de Physique, Universit\'e de Neuch\^atel\\
rue A.-L.~Br\'eguet 1, CH-2000 Neuch\^atel, Switzerland}\\
\smallskip
{\tt frank.ferrari@unine.ch}
\end{center}
\vfill\noindent
This is a set of lectures on the gauge/string duality and non-critical
strings, with a particular emphasis on the discretized, or matrix model,
approach. After a general discussion of various points of view, I describe
the recent generalization to four dimensional non-critical (or five
dimensional critical) string theories of the matrix model approach. This
yields a fully non-perturbative and explicit definition of string theories
with eight (or more) supercharges that are related to four dimensional CFTs
and their relevant deformations. The space-time as well as world-sheet
dimensions of the supersymmetry preserving world-sheet couplings are
obtained. Exact formulas for the central charge of the space-time
supersymmetry algebra as a function of these couplings are calculated. They
include infinite series of string perturbative contributions as well as all
the non-perturbative effects. An important insight on the gauge theory side
is that instantons yield a non-trivial $1/N$ expansion at strong coupling,
and generate open string contributions, in addition to the familiar closed
strings from Feynman diagrams. We indicate various open problems and future
directions of research.

\vfill

\noindent To appear in {\it L'Unit\'e de la Physique fondamentale:
Gravit\'e, Th\'eories de Jauge et Cordes,} 
Les Houches summer school 2001, Session LXXVI.
\medskip
%
%\vfill
%\begin{flushleft}
%February 2002
%\end{flushleft}
%
\newpage\pagestyle{plain}
\baselineskip 16pt
\setcounter{footnote}{0}
%La ligne suivante permet la numerotation des equations par section
%\renewcommand{\theequation}{\thesection.\arabic{equation}}
% La ligne suivante fait pareil mais dans le format amsmath
%\numberwithin{equation}{section}
%\tableofcontents\pagenumbering{arabic}

%
\section{Introduction}

String theory is an unequalled subject for the extensive techniques that it 
uses and the scope of ideas on which it relies. This great variety 
certainly is at the origin of part of the excitement in the field. Yet, 
it pertains to the weakest point of the theory: the lack of 
unifying principles, and of a consistent non-perturbative 
definition on which to rely. It is thus essential to stand back and to
strive to find a synthesis, if only a partial one. It is with this 
motivation that we will discuss below the gauge theory/string theory
duality. We will propose a point of view \cite{fer1} that 
connects different approaches developed over the years. From
this perspective, we are able to obtain several new results, including 
explicit non-perturbative definitions of many string theories
and non-trivial exact formulas.

\section{Many paths to the gauge/string duality}

We start by discussing succinctly some of the many different facts
that suggest a gauge theory/string theory duality.
\subsection{Confinement}
A compelling evidence for the relationship between
ordinary four dimensional 
gauge theories like QCD and string theory is the similarity of their 
particle spectra. In both 
cases one expects to find an infinite set of resonances with masses on a 
Regge trajectory,
\begin{equation}
\label{Regge}
M_{J}^{2} \sim {J\over\alpha '}\, \cvp
\end{equation}
where $M_{J}$ is the mass of the resonance, $J$ its spin,
and $\alpha '$ sets the length 
scales. In the string picture, $1/\alpha '$ is identified with the string 
tension. In the gauge theory picture, the dimensionless coupling constant
$g_{\rm YM}^{2}$ is replaced by a scale $\Lambda$ after dimensional 
transmutation, and $1/\alpha'\sim\Lambda^{2}$. Ordinary four dimensional 
Yang-Mills is believed to confine, and the spectrum (\ref{Regge}) is 
characteristic of a confining gauge theory. Confinement is 
the consequence of the collimation of the chromoelectric flux lines,
which generalize the ordinary Faraday flux lines of electrodynamics. 
The collimation can be demonstrated in the strong coupling 
approximation on the lattice \cite{Wilson}, where it appears to be a 
consequence of the compactness of the gauge group. The string theory dual 
is simply the theory describing the dynamics of the tubes that the 
collimated flux lines form. The collimation of flux lines is interpreted
in terms of a dual superconductor picture \cite{Mandel}. 
The idea is that the relevant degrees of freedom in the strongly coupled 
Yang-Mills theory are magnetically charged. One then assumes that those 
magnetic charges condense, which implies that the chromoelectric 
flux is squeezed into vortices, in the same way as the condensation 
of electric charges squeezes the magnetic field into Abrikosov vortices 
in an ordinary superconductor. 

The chromoelectric flux vortices, or tubes, have a definite thickness
of order $1/\Lambda$. The above description in terms of strings thus
seems to be at best phenomenological. The well-studied fundamental
strings \cite{polbook} indeed have zero thickness. Equivalently, one
would need a fundamental description of relativistic theories based on
light electric and magnetic charges, and none were known until the
proposal in \cite{fer1}. Those difficulties explain why the excitement
initiated in the late sixties eventually faded, and the subject
remained dormant for several decades.
\subsection{Large $N$}
A seemingly very different argument in favor of the gauge 
theory/string theory duality is due to 't Hooft \cite{tHooft}. In an $\suN$ 
gauge theory, with fields transforming in the adjoint representation, 
the Feynman graphs can be depicted using a double-line representation 
corresponding to the double-index notation $A^{m}_{\bar n}$ for the fields. 
This representation makes the relationship with discretized Riemann
surfaces obvious (see Figure 1). Moreover, by taking $g^{2}_{\rm 
YM}N$ (or equivalently after renormalization the mass scale $\Lambda$)
to be an $N$-independent constant, surfaces of genus $g$ comes with a 
factor $1/N^{2g}$. The large $N$ expansion of
$\suN$ Yang-Mills theory is thus a reordering of the Feynman diagrams with
respect to their topology. This shows that a perturbative
discretized closed oriented 
string theory of coupling constant $\kappa\sim 1/N$ is equivalent to the 
reordered perturbative Yang-Mills theory. By perturbative, we mean that the 
correspondence works a priori for the contributions to the Yang-Mills path 
integral that can be represented in terms of Feynman diagrams. Note that 
the world-sheet of the string theory is embedded in the four dimensional 
Minkowski space, since the real-space Feynman rules imply that 
each vertex comes with a space-time label on which we integrate. 

The weakness of the above argument is obvious: the discretized world sheets 
are only vaguely reminiscent of the continuous world-sheets of ordinary 
string theory. Only very large Feynman graphs, which have a very large 
number of faces, may give a good approximation to continuous world sheets.
One could then argue that in the IR, which is the relevant regime for
confinement, the gauge coupling is large and thus the large 
Feynman graphs indeed dominate. This picture is only heuristic and 
might at best yield an effective string theory description, similar to 
the one discussed in Section 2.1. We will see however that it is
extremely fruitful to pursue this idea further. In a slightly different 
context, and with an important additional physical input, we will be able 
to use the 't Hooft representation in a controlled way.

\begin{figure}
\centerline{\epsfig{file=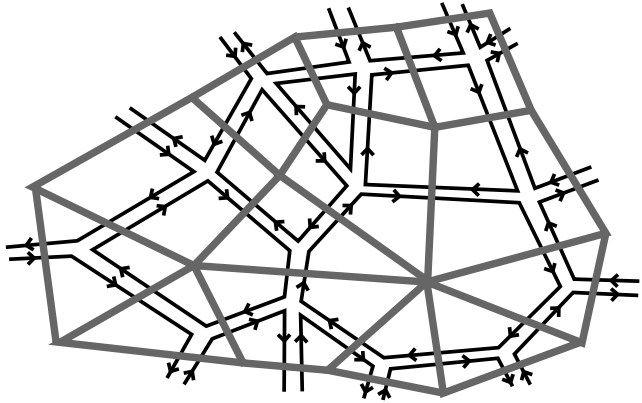,width=8.5cm}}
\caption{Part of a typical Feynman diagram in an $\suN$ gauge 
theory, depicted in the double line representation.
A dual representation (gray lines) is obtained by associating a $p$-gon
to each vertex of order $p$. The $p$-gons generate an oriented
discretized Riemann 
surface. The power of $1/N^{2}$ counts the genus of these surfaces.
\label{mat}}
\end{figure}
\subsection{D-branes}
\begin{figure}
\centerline{\epsfig{file=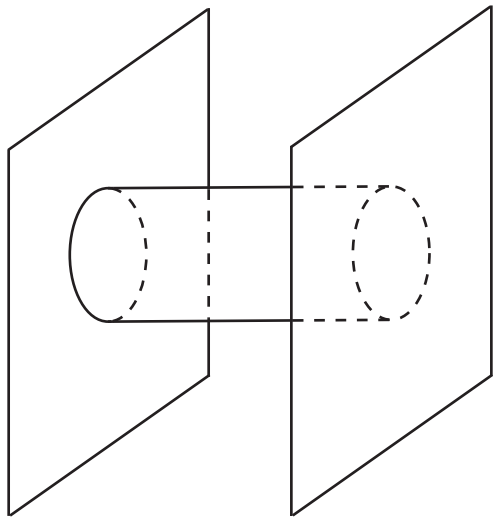,width=6.5cm}}
\caption{Interaction between D-branes. One can interpret the picture 
as either a one-loop diagram for an open string whose ends are stuck 
on the branes (this is described by a Yang-Mills 
theory), or as a tree diagram for a closed string propagating in the 
bulk (this is described by a quantum gravity theory). 
\label{Dbra}}
\end{figure}

Yet another argument in favor of the gauge/string correspondence has 
been through the use of D-branes \cite{malda}. A D$p$-brane can be 
described perturbatively as a $p+1$ dimensional defect in space-time on 
which open strings can end \cite{pol}.
When $N$ D$3$-branes are put on top of 
each other, the low energy dynamics of the open strings is described 
by the ${\cal N}=4$ supersymmetric Yang-Mills theory with gauge 
group $\suN$. On the other hand, at large $N$, the whole of the
$N$ D$3$-branes is an heavy object that can be described by a 
classical solitonic solution in type IIB supergravity \cite{Stro}. 
These complementary descriptions of the D-branes (see Figure 2),
involving either 
open strings or a soliton in a closed string theory, is equivalent to 
the gauge/string duality. This picture can be made precise,
and we refer the reader to other lectures at this school 
\cite{MalKle} in which many details about this correspondence can be 
found.

The above construction fits very well with the discussion of Section 
2.2. The closed string coupling constant indeed turns out to be 
proportional to $1/N$, and $g^{2}_{\rm YM}N$ is independent of this 
coupling. However, 
it seems at first sight impossible to make it consistent with the 
discussion of Section 2.1. The ${\cal N}=4$ theory is 
indeed a conformal field theory, and thus does not confine. There 
is then no length scale to set the string tension. Moreover, 
the correspondence involves type IIB strings, which are of zero 
thickness and live in ten dimensions! Remarkably, it is the very
fact that more than four dimensions are involved that makes the use 
of fundamental strings and the description of conformal gauge theories 
possible. Of the six additional dimensions, five play a rather 
technical and model-dependent r\^ole. 
They are best viewed as additional degrees of freedom 
on the world sheet that are necessary to account for the many ``matter''
fields of the ${\cal N}=4$ super Yang-Mills theory.
Those fields transform under an ${\rm SO}(6)$ R-symmetry, 
which explains why the five dimensions actually form a 5-sphere. The 
other, ``fifth'' dimension $\phi$, is much more interesting.
The form of the non-trivial five dimensional 
metric is actually fixed by conformal invariance to be the AdS$_{5}$ 
metric,
\begin{equation}
\label{met}
\d s^{2}_{5} = \d\phi^{2} +  e^{2\phi /R}\, \d s_{4}^{2}\, ,
\end{equation}
where $\d s_{4}^{2}$ is the four dimensional Minkowski metric. The 
radius $R$ turns out to be proportional to $(g^{2}_{\rm YM}N)^{1/4}$.
From (\ref{met}) we see explicitly that $\phi$ can be interpreted as a 
renormalization group flow parameter, since a shift in $\phi$ can be 
absorbed in a rescaling of the Minkowski metric. The string tension 
varies in the fifth dimension and is set by the RG scale! The 
original open strings that generate the Yang-Mills dynamics are 
naturally attached to the brane in the far UV, $\phi\rightarrow\infty$,
where all the information about the gauge theory is encoded.
Confining theories can be obtained by adding some relevant 
operators to the ${\cal N}=4$ Yang-Mills action. The form of the 
metric is then in general
\begin{equation}
\label{metgen}
\d s^{2}_{5} = \d\phi^{2} +  a^{2}(\phi) \d s_{4}^{2}\, ,
\end{equation}
where the function $a(\phi)$ is well approximated by $\exp (\phi 
/R)$ in the UV region $\phi\rightarrow\infty$, but remains to be 
determined in the IR. At the confining scale 
$\phi=\phi_{*}\sim\ln\Lambda$, $a(\phi)$ will typically have a minimum. 
It is then energetically favoured for the fundamental strings to sit 
in this region, and this implies that they acquire an effective
thickness $1/\Lambda$ from the four dimensional point of view \cite{thick}.

The three aspects discussed in Section 2.1, 
2.2 and 2.3 are thus fully consistent, in a subtle and interesting way.
We will better understand 
the origin of the fifth dimension in the next subsection, but
the fact that conformal field theories play a prominent r\^ole 
remains a rather weird and not very well understood feature. 
The discussion in Section 3 will shed some light on
this part of the story.
\subsection{Non-critical strings}
The most direct approach to construct a string theory dual to a four 
dimensional field theory is to try to quantize directly the string 
with a target space of dimension $D=4$. Because of the quantum 
anomaly in the Weyl symmetry,
the world sheet metric $g_{ab}$, which classically is a mere auxiliary
field, does not decouple. Using world sheet 
diffeomorphisms, we can always put the metric in the form
$g=\hat g e^{\phi}$, where $\hat g$ depends on a finite number 
of moduli. The conformal factor $\phi$, which is called the Liouville 
field, then acquires a non-trivial
world sheet dynamics. When $D\leq 1$, or more precisely for a world-sheet 
theory of central charge $c\leq 1$, it is possible, under rather 
well-controlled assumptions, to work out this dynamics explicitly 
\cite{pola}. 
For example, in the case $D=1$, the classical world sheet action is
\begin{equation}
\label{wsaD1}
S = {1\over 4\pi\alpha'}\int\! \d^{2}\sigma\sqrt{g}\left(
\partial_{a} X\partial^{a} X +\lambda_{0} + \alpha' \Phi R\right)\, .
\end{equation}
The world sheet scalar $X$ corresponds to the embedding coordinate and
the string coupling constant is $g_{{\rm s}0} = e^{\Phi}$.
A world sheet cosmological constant $\lambda_{0}$ must also be included 
because the Weyl symmetry is broken quantum 
mechanically. Integrating out $g$ then yields an action of the form
\begin{equation}
\label{wsaq1}
S = {1\over 4\pi\alpha'}\int\! \d^{2}\sigma\sqrt{\hat g}\left( 
\partial_{a}X\partial^{a}X + \partial_{a}\phi\partial^{a}\phi + T(\phi)+
\alpha'\Phi (\phi) \hat R \vphantom{g^{ab}\partial_{a}}\right)\, ,
\end{equation}
which shows that the Liouville field $\phi$
plays the r\^ole of a new dimension. The physics 
is not uniform in this new dimension because of the non-trivial 
``tachyon'' $T$ and dilaton $\Phi$ backgrounds, which are 
determined by requiring world sheet conformal invariance. More 
details may be found in the review \cite{revL}. Unfortunately, the case
$D>1$ is much 
more difficult and a direct analysis has never been performed. 
However, it can be noted \cite{polb} that the most general ans\"atz 
compatible with the symmetries of the problem, in particular with 
$D$-dimensional Lorentz invariance, is
\begin{equation}
\label{wsaq2}
S = {1\over 4\pi\alpha'}\int\! \d^{2}\sigma\sqrt{\hat g}\left( 
a^{2}(\phi)\partial_{a}X_{\mu}\partial^{a}X^{\mu} + 
\partial_{a}\phi\partial^{a}\phi + \cdots
\vphantom{g^{ab}\partial_{a}}\right)\, ,
\end{equation}
where the dots represent various possible background fields. The main 
difference with the $D\leq 1$ case is that the $D+1$ dimensional metric 
is not flat. When $D=4$, the Liouville field is but the fifth 
dimension discussed in Section 2.3, and we recover in particular 
equation (\ref{metgen}). For the string theory (\ref{wsaq2}) to 
describe a gauge theory, additional consistency conditions must be 
satisfied. In particular, the open strings generating the Yang-Mills 
dynamics must live either at the horizon $a=0$ or at the point of 
infinite string tension $a=\infty$. The latter choice, that 
corresponds to a ``fundamental'' brane living in the far UV, seems more 
natural and is consistent with the discussion of Section 2.3.
In either case, the 
open string theory will have only vector states and the zig-zag 
symmetry of the Wilson loops will be satisfied \cite{polb}.

String theories based on actions like (\ref{wsaq2}) are extremely difficult
to solve. Even in the simplest case $D\leq 1$ described by (\ref{wsaq1}), 
only a partial analysis can be given. The reason is twofold: first 
the world sheet theories are complicated interacting two dimensional field 
theories, which makes the analysis of string perturbation theory 
particularly involved; second the string coupling can grow due to the 
non-trivial dilaton background, which can altogether 
invalidate the use of the string perturbative framework. In spite of 
these daunting difficulties, the case $D\leq 1$ has actually been solved
to all orders of string perturbation theory in a series of remarkable
papers \cite{k0,k1,k2,k3}. The basic idea \cite{k1} is to 
consider a discretized version of the $D$ dimensional
string theory, which 
turns out quite surprisingly to be easier to study than the 
original continuum model. The continuous world 
sheets are approximated by discretized surfaces made up of flat 
polygons of area $\ell^{2}$. The curvature is concentrated on a 
discrete lattice $L$ on which the vertices of the polygons lie. The 
discrete
world sheet fields are defined at the center of the polygons, or 
equivalently on the dual lattice $\tilde L$.
For example, the discrete Polyakov path integral $Z_{\rm P}(\ell)$ defining
string perturbation theory based on the action (\ref{wsaD1}) with the 
cut-off $1/\ell$ can be written
\vfill\eject
%(the index P could be for Polyakov or Perturbative)
%
\begin{eqnarray}
&&\hskip -1cm Z_{\rm P}(a)=\sum_{h\geq 0}
\int\limits_{\scriptstyle {\rm genus\ } 
h\atop\scriptstyle\rm world\ sheets} \hskip -.6cm
[\d g_{ab}(\sigma) \d X(\sigma)]\, e^{-S}\,{\lower 
10pt\hbox{\strut\vrule}}_{\, {\rm cut-off\ } 1/\ell} = \nonumber \\
&&\hskip 2cm\sum_{h\geq 0}g_{{\rm s}0}^{2h-2}\hskip -.3cm
\sum_{\scriptstyle {\rm genus\ }h\atop \scriptstyle{\rm lattices\ }L} 
\!\!\! e^{-{\lambda_{0} \ell^{2}|\tilde L|\over 4\pi\alpha'}}
\int\!\prod_{i\in\tilde L} \d X_{i}\!\!\!\prod_{\scriptstyle
{\rm links\ }\langle kl\rangle\atop\scriptstyle {\rm of\ }\tilde L}
\!\!\!\!\Delta \left(|X_{k}-X_{l}|/\smash{\sqrt{\alpha'}}\strut\right)\, .
\label{dispolint}
\end{eqnarray}
In the above formula, $|\tilde L|$ represents the number of vertices of
the lattice $\tilde L$, or equivalently the number of polygons in the
discretization of the world sheet.
The function $\Delta$ is a gaussian and is derived from the kinetic term 
for $X$ in (\ref{wsaD1}). The cosmological constant $\lambda_{0}$ can 
a priori be renormalized, since the sum over lattices of fixed size 
and the integration over the $X_{i}$s can generate a counterterm 
proportional to $|\tilde L|$.

We can now use 't~Hooft's idea described in Section 2.2.
The very involved combinatorial 
problem corresponding to the discrete lattice sum (\ref{dispolint}) is 
conveniently encoded in the Feynman graph expansion 
of a matrix theory. Originally, 't Hooft considered four dimensional
Yang-Mills theories, but the argument is straightforwardly extended to any 
matrix field theory. If the matrix field theory lives in $D$ 
dimension(s), then the discretized world sheets are embedded in the 
$D$ dimensional Minkowski space. It is easy to see that the matrix 
theory corresponding to (\ref{dispolint}) is a quantum 
mechanics based on a single $N\times N$ hermitian matrix $M$,
and that we have
\begin{equation}
\label{MatQM}
e^{Z_{\rm P}(\ell)} = 
\int\! \left[\smash{\d^{N^{2}}}\! M(\tau)\strut\right]\,
\exp\left[ -N\int\!\d\tau\, \tr\Bigl( {1\over 2} \dot M^{2} + 
{1\over 2\alpha'} M^{2} + U(M)\Bigr)\right] .
\end{equation}
The euclidean time $\tau$ corresponds 
to the embedding coordinate $X$ in the string theory. Terms in 
$M^{p}$ in the potential $U(M)$ generate $p$-gons in the 
discretization of the world sheets, see Figure 1. The metric on the 
world sheet is determined by giving an area $\ell^{2}$ to the $p$-gons.
We can actually restrict ourselves to the simple potential
\begin{equation}
\label{U1}
U(M) = {g\over 3!} M^{3}\, .
\end{equation}
More complicated potentials yield the 
same theory in the continuum limit. In terms of 
the matrix theory variables, we have
\begin{equation}
\label{Zpolmat}
Z_{\rm P}(\ell) = \sum_{h\geq 0} N^{2-2h}\hskip -.3cm
\sum_{\scriptstyle {\rm genus\ }h\atop \scriptstyle{\rm lattices\ 
}\tilde L} \!\!\! (-g)^{|\tilde L|}
\int\!\prod_{i\in\tilde L} \d\tau_{i}\!\!\!\prod_{\scriptstyle
{\rm links\ }\langle kl\rangle\atop\scriptstyle {\rm of\ }\tilde L}
\!\!\!\!\Delta_{\rm E} \left(|\tau_{k}-\tau_{l}|/\smash{\sqrt{\alpha'}}
\strut\right)\, .
\end{equation}
The power of $N$ comes from the standard 't Hooft's analysis, and 
yields a string coupling $g_{{\rm s}0}=1/N$. The power
$|\tilde L|$ of $g$ corresponds to the number of vertices in the ordinary
representation of the Feynman diagrams, or equivalently to the number of
polygons in the dual representation, see Figure 1. The sum is over the
lattices $\tilde L$, which is of course equivalent to the sum over the dual
lattices $L$ as in (\ref{dispolint}). 

The matrix quantum mechanics
(\ref{MatQM}) can be mapped onto a problem of free fermions \cite{k0},
and is thus exactly solvable. In particular, the discrete Polyakov
partition function, which is proportional to the radius $r$ of the
embedding dimension due to translational invariance, can be calculated 
since it is related to the ground state energy $E(g)$,
\begin{equation}
\label{ZE}
E(g) = -\lim_{r\rightarrow\infty} {Z_{\rm P}(\ell)\over r}\,\cdotp
\end{equation}
Any other observable of the discrete string theory could be related to 
quantum mechanical amplitudes in a similar way.
However, and as was stressed in Section 2.2, we are still far from our 
goal of solving a continuum string theory. Indeed, we must find a
way to implement
consistently the continuum limit $\ell\rightarrow 0$. At finite $\ell$,
the identification of (\ref{dispolint}) and 
(\ref{Zpolmat}) must be taken with a grain of salt, because
the link factor $\Delta_{\rm E}$ is the 
one dimensional euclidean propagator, which is a simple exponential, 
unlike the gaussian $\Delta$ that corresponds to the Polyakov action. 
On the other hand, it is very plausible that 
any link factor for which a consistent continuum limit can be 
defined will yield the unique consistent continuum string theory.

The fact that a continuum limit can be defined for (\ref{Zpolmat}) 
relies on the non-trivial property that the average number of polygons 
in the relevant Feynman graphs for (\ref{MatQM}) diverges for a negative
critical value of the coupling $g$ \cite{k0,k1}. Indeed,
the ground state energy admits an expansion
\begin{equation}
\label{Eexp}
E(g) = \sum_{h\geq 0} N^{2-2h} E_{h}(g) =
\sum_{h\geq 0} g_{{\rm s}0}^{2h-2} E_{h}(g)\, ,
\end{equation}
where
\begin{equation}
\label{Ehexp}
E_{h}(g) = \sum_{k=0}^{\infty} E_{h,k} (-g)^{k}\, .
\end{equation}
The numerical coefficient $E_{h,k}<0$ gives the contribution from 
genus $h$ surfaces with a fixed number $k$ of polygons. It is given by 
the terms in (\ref{Zpolmat}) with $|\tilde L|=k$.  
The series (\ref{Ehexp}) has a finite radius of convergence which is
independent of $h$, and the critical coupling corresponds to 
the point where it diverges \cite{k0}. This means that the large $N$ 
expansion of the matrix quantum mechanics breaks down at $g=g_{\rm c}$.
When $g\rightarrow g_{\rm c}$, 
the terms with a high power of $k$, or equivalently the surfaces with 
a large number of polygons, dominate the sum (\ref{Ehexp}). {\it The limit 
$g\rightarrow g_{\rm c}$ is thus a continuum limit in which
't~Hooft's heuristic picture of Section 2.2 becomes precise.} At large 
$k$, $E_{h,k}$ picks up a term proportional to $(-g_{\rm c})^{-k}$,
and thus we expect the
contribution of surfaces of size $|\tilde L|$ to be proportional to 
$(g/g_{\rm c})^{|\tilde L|}\simeq \exp \left(-\smash{|\tilde L|(g_{\rm 
c}-g)}\strut\right)$. There is actually a logarithmic 
correction to that formula, and $g_{\rm c}-g$ is replaced by
$\epsilon$ defined by
\begin{equation}
\label{egrel}
g_{\rm c}-g = -\epsilon\ln\epsilon\, .
\end{equation}
Comparing with (\ref{dispolint}), we 
get the renormalized world sheet cosmological constant
\begin{equation}
\label{ccren}
\lambda = 4\pi\alpha' {\epsilon\over \ell^{2}}\,\cdotp
\end{equation}
This equation gives the precise relation between the coupling $g$ of the 
matrix quantum mechanics and the world sheet cut-off $1/\ell$. In the 
continuum limit, (\ref{Eexp}) na\"\i vely suggests that $E(g)$ 
diverges, since the fixed genus contributions $E_{h}(g)$ themselves
diverge. However, the limit can still be made consistent because 
the divergences are very specific, and can be compensated for by a 
simple multiplicative renormalization of the string coupling $g_{{\rm 
s}0}$ \cite{k1,k2,k3}. Indeed, we have
\begin{equation}
\label{Ehas}
E_{h}(g) \mathop{\propto}\limits _{g\rightarrow g_{\rm c}} 
{1\over\epsilon^{2h-2}}\,\cvp
\end{equation}
which shows that in the double scaling limit \cite{k2,k3}
\begin{equation}
\label{dsca1}
N\rightarrow\infty\, ,\quad \epsilon\rightarrow 0\, ,\quad 
N\epsilon = {\rm constant} = 1/\kappa\, ,
\end{equation}
the ground state energy (\ref{Eexp}) has a finite limit $E_{\rm 
scaled}$, with an expansion of the form\footnote{The sphere and torus 
contributions are actually logarithmically divergent. This can be 
understood from the point of view of the continuum theory, but is 
beyond the scope of the present review. Details can be found in the 
excellent lecture notes \cite{klebrev}.} 
\begin{equation}
\label{Eexp2}
E_{\rm scaled} = \sum_{h\geq 0} E_{h}\,\kappa^{2h-2}\, .
\end{equation}
The double scaling limit (\ref{dsca1}) shows that the renormalized
string coupling is
\begin{equation}
\label{scren}
g_{\rm s} = {1\over N\ell^{2}}= {g_{{\rm s}0}\over \ell^{2}}\, \cvp
\end{equation}
and thus have a non-trivial world sheet dimension two. The 
dimensionless coupling $\kappa$ is a combination of the genuine 
string coupling and of the cosmological constant,
\begin{equation}
\label{kappadef}
\kappa = {4\pi\alpha' g_{\rm s}\over\lambda}\,\cdotp
\end{equation}

The non-critical strings based on a world sheet theory 
of central charge $c<1$ can be treated in a similar way. The 
matrix quantum mechanics (\ref{MatQM}) is replaced by a simple matrix 
integral
\begin{equation}
\label{Matint}
\int\! \d^{N^{2}}\! M\, e^{-N \tr U(M;g_{j})}
\end{equation}
with a general potential
\begin{equation}
\label{Ugen}
U(M;g_{j}) = {1\over 2} M^{2} + \sum_{j=3}^{p} {g_{j}\over j!} M^{j}\, .
\end{equation}
By adjusting $k$ of the couplings to special values, we can go to a 
$k^{\rm th}$-order critical point, called the $k^{\rm th}$ Kazakov 
critical point, for which $c=1-3(2k-3)^{2}/(2k-1)$. The double scaling 
limit (\ref{dsca1}) takes the general form
\begin{equation}
\label{dsca2}
N\rightarrow\infty\, ,\quad \epsilon\rightarrow 0\, ,\quad 
N\epsilon^{1-\gamma_{\rm str}/2} = {\rm constant}\, ,
\end{equation}
with some exponent $\gamma_{\rm str} = -1/(k+1)$, but qualitatively 
the results are very similar to the $c=1$ case.
All the $(p,q)$ minimal CFTs can 
be obtained in this way, by considering a slight generalization of 
(\ref{Matint}) based on a two-matrix model. All the correlators
can be studied, and the RG flows between the various theories 
are described by a nice mathematical structure based on the KdV 
hierarchy. Details can be found in \cite{revZJ}.
\vfill

\begin{table}[h]\centerline{
\begin{tabular}{|m{2.5in}|m{2.5in}|}
\hline
\vbox to 1.75 in{\vskip .25in
\hbox to 2.5in{Continuous world sheets with $c\leq 1$}\vskip .1in
\hbox to 2.5in{\hfill $\displaystyle Z=\int [\d g_{ab}(\sigma)\d 
X(\sigma)]\, e^{-S}$\hfill}\vfill
\hbox to 2.5in{\hfill $c+1$ dimensional flat target space\hfill}
\vskip .1in
\hbox to 2.5in{\hfill $\d s^{2} = \d\phi^{2}+\d 
X^{2}$\hfill}} &\vbox to 1.75in {\vskip .25in
\hbox to 2.5 in{Discretized world sheets with $c\leq 1$}\vskip .1in
\hbox to 2.5 in {\hfill$\displaystyle e^{Z(\ell)}=\int\! \d^{N^{2}}\! M\,
e^{-N \tr U(M;g_{j})}$\hfill}\vfill
Matrix integral with Kazakov critical points in the double 
scaling limit\vskip .03in}\\ 
\hline
\vbox to 1.75in{\vskip .25in\hbox to 2.5 in{Continuous
world sheets with $c> 1$}\vskip .1in 
\hbox to 2.5in{\hfill $\displaystyle Z=\int [\d g_{ab}(\sigma)\d 
\vec X(\sigma)]\, e^{-S}$\hfill}\vfill
\hbox to 2.6in{\hfill $c+1$ dimensional curved target space\hfill}
\vskip .1in
\hbox to 2.5 in{\hfill $\d s^{2} = \d 
\phi^{2} + a^{2}(\phi)\, \d\vec X^{2}$\hfill}} &
\vbox to 1.75in {\vskip .25in\hbox to 2.5in{Discretized world sheets
with $c> 1$}
\vfill\hbox to 2.5 in{\hfill\HUGE ?\hfill}\vfill} \\ \hline
\end{tabular}}
\caption{The magic square of the non-critical string theories. The
continuum approach side of the $c=1$ 
barrier was finally cleared in 1997, but the discretized approach side 
remained unviolated until recently. We will try to shed some light on the 
material hidden behind the question mark in the following.}
\end{table}
\eject

Let us conclude this introduction with an important remark. The simple
potential (\ref{U1}) does not have a ground state. This means that
$E(g)$ can only be defined as a sum over Feynman graphs. We could try
to use another potential, for example $U(M) = gM^{4}/4!$, but the
positivity of the partition function (\ref{dispolint}) implies that
the coupling $g$ in (\ref{Zpolmat}) must be negative. As explained in
details in Section 7 of the review \cite{revZJ}, it is actually
impossible to obtain a non-perturbative definition of the unitary
string theories using the matrix models. The ``solutions'' of the
$D\leq 1$ string theories thus yield the observables to all orders of
perturbation theory, but not beyond. Of course, the Polyakov
formulation of string theory, that was our starting point, is
perturbative in nature. Equations like (\ref{ZE}) are thus perfectly
consistent and must be understood as a statement about the asymptotic
perturbative series. However, it was originally expected that the
powerful new formulation in terms of a matrix theory in the double
scaling limit could give insights into a non-perturbative definition
of string theory. The fact that this is not the case is a major
drawback of the classic matrix model approach. We will have much more
to say in Section 3 about this very important point of principle.

\section{Four dimensional non-critical strings}

The discretized approach advocated in \cite{k1} is indisputably
the most fruitful to study the $D\leq 1$ string theories (\ref{wsaq1}).
It is then most natural to try to extend the same ideas to the 
$D>1$ theories (\ref{wsaq2}). The first concrete results in 
this direction were obtained only recently by the present author 
\cite{fer1}. The results of \cite{fer1}, together 
with the insights gained in a series of preparatory \cite{fer0, fer0b} and 
subsequent \cite{fer2,fer3} works, strongly suggest that the 
discretized approach in $D>1$ (and in particular in the case $D=4$ 
on which we will focus) has considerable power. In particular, it 
provides an explicit {\it non-perturbative} definition of the $D=4$ 
non-critical strings, in sharp contrast with the $D\leq 1$ cases.

The first step to go from $D\leq 1$ to $D=4$ is straightforward: the $D=0$ 
matrix integrals (\ref{Matint}) or $D=1$ matrix path integral (\ref{MatQM})
are replaced by $D=4$ matrix path integrals
\begin{equation}
\label{Mat4d}
\int\! \left[ \d M(x)\strut\right]\,
\exp\left[ -N\int\!\d^{4}x\, {\cal L}(M,\partial M)\right] ,
\end{equation}
where $M$ represents in general a collection of $N\times N$ hermitian 
matrices and $\cal L$ is a lagrangian density. In the original 't~Hooft's 
example \cite{tHooft}, we have four matrices corresponding to the four 
components of the vector potential $A_{\mu}$, and $\cal L$ is 
the $\suN$ Yang-Mills lagrangian with the 't Hooft coupling 
$g_{\rm YM}^{2} N$ or equivalently after renormalization
the dynamically generated scale
$\Lambda$ chosen to be independent of $N$.
The path integral (\ref{Mat4d}) defines a 
discretized string theory with a four dimensional target 
space.\footnote{Considering $\suN$ instead of ${\rm 
U}(N)$ amounts to integrating over traceless hermitian matrices in 
(\ref{Mat4d}). This does not change the discretized string 
interpretation in the large $N$ limit.} Of course, and
as stressed in Section 2.2, the real challenge is to succeed in
taking the continuum limit. This was done in Section 2.4 for the simple 
$D\leq 1$ integrals (\ref{MatQM}) or (\ref{Matint}) by adjusting the 
parameters in the interaction potential to special Kazakov critical
points. However, the pure Yang-Mills path integral does not have any
free parameter, and thus this procedure cannot be straightforwardly
extended.

The way out of this problem was first proposed in \cite{fer0}. The 
idea is to consider gauge theories with Higgs fields in the adjoint 
representation of the gauge group. The adjoint Higgs fields correspond 
to additional hermitian matrices $M$ on which we integrate in
(\ref{Mat4d}). The Higgs theory depends on couplings parametrizing 
the Higgs potential. Varying those couplings amounts to varying the 
masses of the gauge bosons, or equivalently the effective gauge 
coupling. It was then argued in \cite{fer0} that for some special values of the Higgs
couplings, that correspond to W bosons of masses of order $\Lambda$,
critical points may exist. It could then be possible to define a
consistent continuum limit, in strict parallel to the $D\leq 1$ cases.
The consistency of those ideas were checked on a simple
toy model \cite{fer0}. Supersymmetry plays no r\^ole in the 
discussion, and in particular the model studied in \cite{fer0} was 
purely bosonic. Unfortunately, and even though lattice calculations 
seem encouraging \cite{lat}, the present-day analytical tools do not 
make the search for critical points in purely bosonic 
Yang-Mills/adjoint Higgs theories possible.

This is where supersymmetry enters the game: we do have some control on 
supersymmetric Yang-Mills/Higgs theories, and we can try to apply our 
ideas in this context. A typical example is pure ${\cal N}=2$ 
supersymmetric $\suN$ gauge theory. The adjoint 
Higgs field is automatically 
included in this theory because it is a supersymmetric partner
of the gauge bosons. Strictly speaking, this 
theory is parameter-free, because the Higgs coupling is related 
to the gauge coupling by supersymmetry, and the latter is replaced by a 
mass scale in the quantum theory. However, there is a freedom in 
the choice of the Higgs expectation values, because the Higgs potential 
has flat directions that are protected by a non-renormalization theorem.
The path integral (\ref{Mat4d}) is then parametrized 
by a set of boundary conditions for the Higgs field at infinity.
Such parameters are called moduli of the field theory, and
one can show that there are $N-1$ moduli for the pure $\suN$ theory. 
For our purposes, the moduli will play the r\^ole of the couplings
$g_{j}$ in the potential (\ref{Ugen}) of the 
$D\leq 1$ theories. In more general ${\cal N}=2$ gauge 
theories, one may have quark mass parameters in addition to the 
moduli. Mass parameters and moduli are on an equal footing in our 
discussion, and we will denote them collectively by ${\cal M} =\{m_{j}\}$.

There is a subtle but fundamental difference between the space of 
couplings $g_{j}$ for the simple matrix integrals of Section 2 and the 
space $\cal M$ for the gauge theories: the Yang-Mills path 
integrals are non-perturbatively defined for {\it all} values of the 
moduli or parameters, unlike the matrix integrals in $D\leq 1$. As 
discussed at the end of Section 2.4, this has some important 
consequences, because the Kazakov critical 
points in $D\leq 1$ always lie at the ``wrong'' couplings, for example
$g<0$ for the integral (\ref{MatQM}) with a quartic potential.
As a consequence, the $D\leq 1$
continuum string theories can only be defined in perturbation theory. 
We see that in four dimensions, we cannot run into this problem: we 
can only get {\it non-perturbative} definitions of string
theories, from the {\it non-perturbative} gauge theory path integrals. 
The hard part is of course to find Kazakov critical points on 
$\cal M$.
\subsection{Four dimensional CFTs as Kazakov critical points}
It is well-known that there are so-called singularities on the 
moduli/parameter space $\cal M$ of ${\cal N}=2$ gauge theories where
a non-trivial low energy physics develops \cite{SW}.
For example, in the pure $\suN$ case, one can adjust
$p\geq 2$ moduli to special values and get an interacting
CFT in the infrared \cite{AD}. The nature of such CFTs has
remained rather mysterious, because the light degrees of freedom 
include both electric and magnetic charges, and thus a conventional
local field theoretic description does not exist. Such theories are 
nevertheless of primary interest, because effective theories of light 
electric and magnetic charges are believed to play an important r\^ole
in real world QCD, as explained at the end of Section 2.1.

The main result of our investigations is then the following:

\noindent {\it The 
non-trivial CFTs on the moduli space of supersymmetric gauge theories, 
that can be obtained by adjusting a finite, $N$-independent, number of 
moduli, are the four dimensional generalizations of the Kazakov critical 
points. They can be used to define double scaling limits that yield 
string theories dual to the corresponding CFT with the possible 
relevant deformations.}

A direct consequence of this claim is that 
the theories of light electric and magnetic charges can be described 
at a fundamental level by string theories, a result in perfect 
harmony with the discussion in Section 2. Moreover, many CFTs, 
including gauge CFTs, can be constructed in this way. For example, 
${\cal N}=4$ super Yang-Mills with gauge group ${\rm SU}(p+1)$ can be 
obtained on the moduli space of a parent ${\cal N}=4$ theory with 
gauge group $\suN$, by adjusting $p$ moduli. The very possibility of
defining 
a double scaling limit automatically demonstrates that there is a 
string dual, whose continuous world sheets are constructed from the
Feynman diagrams of the {\it parent} gauge theory.

A basic property of a Kazakov critical point is that the large $N$ 
expansion of the matrix theory breaks down in its vicinity. This 
is due to divergences at each order in 
$1/N$. For example, the divergence of the coefficient $E_{h}$ 
defined in (\ref{Eexp}) is given by (\ref{Ehas}). 
Those divergences are crucial, since the 
whole idea of the double scaling limits rely on the possibility of 
compensating the divergences by 
taking $N\rightarrow\infty$ and approaching the critical point in a 
correlated way. This procedure picks up the most divergent, universal, 
terms that correspond to the continuum string theory. A basic 
consequence of our claim is thus that the large $N$ expansion of 
${\cal N}=2$ supersymmetric gauge theories should break down at 
singularities on the moduli space! More precisely, one can distinguish 
two classes of singularities: those that are obtained by adjusting a 
large number of order $N$ of moduli, and others that are obtained by 
adjusting a finite, $N$-independent number of moduli. Our claim 
implies that divergences in the large $N$ expansion must be found in 
the second case. This is by itself a rather strong and new statement
about the behaviour of the large $N$ expansion of certain gauge 
theories. If the ${\cal N}=2$ parent gauge theory admits a string dual
(not to be confused with the string theories produced in the double 
scaling limits!), it implies that the string theory does not admit a 
well-defined perturbation theory at the critical points.

The breakdown of the $1/N$ expansion can be given a simple interpretation.
Commonly, trying to find a good approximation scheme to describe a 
non-trivial critical point is difficult.
A typical example is 
$\phi^{4}$ theory in dimension $D$. The theory has two parameters, the 
bare mass $m$ (or ``temperature'') and the bare coupling constant $g$. By 
adjusting the temperature, we can go to a point where we have massless 
degrees of freedom, and a non-trivial Ising CFT in the IR.
The difficulty is that the renormalized fixed point coupling $g_{*}$
is large, and thus ordinary perturbation theory in $g$ fails. It is 
meaningless to try to calculate universal quantities like critical 
exponents as power series in $g$, since those are $g$-independent. 
Either the tree-level, $g$-independent contributions are exact and the 
corrections vanish (this occurs above the critical dimension, which is
$D_{\rm c}=4$ for the Ising model, and we have a trivial
fixed point $g_{*}=0$ well described by 
mean field theory), or the expansion parameter corresponds to a relevant 
operator and corrections to mean field theory are plagued by untamable IR 
divergencies. The critical points on the moduli space of ${\cal N}=2$ 
super Yang-Mills are very similar to the Ising critical point below the
critical dimension,
and $1/N$ is very similar to the $\phi^{4}$ coupling $g$. They are
characterized by a set of critical exponents \cite{AD}
that are independent of the number of colours $N$ of the parent gauge 
theory in which the CFT is embedded. These critical exponents cannot 
consistently be calculated in a $1/N$ expansion.
Even the simple monopole critical points that are known to be trivial
are not described consistently by 
the $N\rightarrow\infty$ limit of the original gauge theory,
because electric-magnetic duality is not 
implemented naturally in this approximation scheme. 
Note that the argument does not apply to CFTs obtained by adjusting a 
large number $\sim N$ of moduli, because those are
$N$-dependent, and the large $N$ expansion can then
certainly be consistent.

The fact that the divergences have an IR origin implies
that the string theories obtained in the double 
scaling limits are dual to the relevant deformations of the CFT at the 
critical point. Indeed, the scaling limits pick up the most IR divergent 
contributions, which are due to the light degrees of freedom only.
This can be checked explicitly \cite{fer1,fer2}, as we will see in 
Section 3.4.

Our arguments so far have been qualitative, and the reader may feel 
rather uncomfortable. Indeed,
it is at first sight hard to imagine how to test our ideas by explicit 
calculations \cite{polrem}. For example, one would 
like to compute explicitly some observables in the $1/N$ expansion, and 
check explicitly that the large $N$ 
expansion does break down at the critical point, and that the 
divergences are specific enough for a consistent double scaling limit 
to be defined. We explain below how such calculations can actually
be done. This involves a rather surprising result about the large $N$ 
behaviour of instanton contributions at strong coupling.
\subsection{Instantons and large $N$}
Unlike the $D\leq 1$ matrix models, which are exactly solvable, only a 
small number of observables can be calculated in ${\cal N}=2$ gauge 
theories, following Seiberg and Witten \cite{SW,Oz}.
Those observables are physically very important, since they 
correspond to the leading term in a derivative expansion of the low energy 
effective action. In particular, the central charge of the supersymmetry 
algebra, and thus the mass of the BPS states, can be calculated 
exactly by using Gauss' law. 
However, those observables are very special mathematically, because they 
pick up only a one-loop term from ordinary perturbation theory. The 
non-trivial physics comes entirely from an infinite series of instanton 
contributions.

Relevant contributions in the large $N$ limit are 
usually assumed to come from the sum of the Feynman graphs at each 
order in $1/N^{2}$ \cite{tHooft}, as reviewed in Section 2. On the 
other hand, instantons are usually disregarded \cite{Witins}. This is 
due to the fact that the 
instanton action is proportional to $N$ in the 't Hooft's scaling
$g_{\rm YM}^{2}\propto 1/N$. The effects of
instantons of topological charge $k$ and size $1/v$ are thus proportional,
in the one-loop approximation which is exact for ${\cal N}=2$ super
Yang-Mills, to
\begin{equation}
\label{instcont}
e^{-8 \pi^2 k/g_{\rm YM}^2} = \Bigl( {\Lambda\over v}\Bigr)^{kN\beta_0}\cvp
\end{equation}
where $\beta_0$ is a coefficient of order $1$ given by the one-loop
$\beta$ function. This formula suggests
that the only smooth limit of instanton contributions when
$N\rightarrow\infty$ is zero, with exponentially suppressed corrections
\cite{Witins}. Large instantons (small $v$),
if relevant, would produce catastrophic exponentially large contributions,
and if one is willing to assume that the large $N$ limit makes sense
the only physically sensible conclusion seems to be that instantons
are irrelevant variables. This argument is independent of 
supersymmetry. In real-world QCD, the ratio $\Lambda /v$ in (\ref{instcont})
would simply be replaced by a more complicated function.
In QCD, instantons of all sizes (all $v$) can
potentially contribute, and this led Witten to argue that the instanton gas
must vanish \cite{Witins}. In Higgs theories, like ${\cal N}=2$ super
Yang-Mills, the Higgs vevs introduce a natural cutoff $v$ on the size of
instantons, and for $v$ large enough (``weak coupling'') the instanton gas
can exist but is just negligible at large $N$. At small $v$ (``strong
coupling''), we run into the same difficulties as in QCD.

The above argument would seem to imply, at least superficially, that the
Seiberg-Witten observables are not 
suited for testing our ideas on the non-trivial behaviour of the large 
$N$ expansion. However, it was realized in \cite{fer0b} that there is 
a major loophole in the analysis based on (\ref{instcont}). The point 
is that (\ref{instcont}) gives the contribution at fixed 
instanton number $k$, whereas the 
full instanton contribution is given by an infinite series of the form
\begin{equation}
\label{insser}
S(v/\Lambda)=\sum_{k=1}^{\infty}c_{k}\Bigl({\Lambda\over 
v}\Bigr)^{kN\beta_{0}}.
\end{equation}
As long as $|v|\gg |\Lambda|$, the series converges, and a term-by-term 
analysis makes sense. In particular, we do expect the
sum $S$ to vanish exponentially at large $N$. However, at strong 
coupling $|v|\ll |\Lambda|$, things are entirely different
because the large instantons make the sum diverge, and it is 
meaningless to isolate a particular term in the sum.
This corresponds to the 
intuitive idea \cite{Witins} that the instanton gas disappears. 
However, this does not imply that $S(v/\Lambda)$ itself is ill-defined
in the strong coupling regime, because the series (\ref{instcont}) can
have a smooth analytic continuation. In some sense, instanton will
transmute into something new through the process 
of analytic continuation. What this might be can be discovered by 
working out explicitly the large $N$ limit of the analytic continuation
of (\ref{insser}) \cite{fer0b},
and the result turns out to be extremely interesting.
\subsection{A toy model example}
Since we do not want to assume that the reader is familiar with the 
technology of the exact results in ${\cal N}=2$ gauge theories, we will 
discuss a simple toy example instead.
Our toy example actually plays a r\^ole in the full calculation in the 
gauge theory \cite{fer0b,fer1}, but it is much simpler and nonetheless 
illustrates the main points 
discussed in Sections 3.1 and 3.2. The model is based on the equation
\begin{equation}
\label{toyex}
P(\sigma)=\sigma\prod_{i=1}^{N-1} (\sigma + v_{i} ) = \Lambda^{N}.
\end{equation}
The observables are the roots of this 
equation. The moduli space $\cal M$ is parametrized by the $v_{i}$s.
Critical points ${\rm C}_{p}$ are obtained when $p\geq 2$ roots 
coincide. The vicinity of ${\rm C}_{p}$ is described by the 
equation
\begin{equation}
\label{reldef}
\sigma^{p} + \sum_{k=2}^{p}t_{p-k}\sigma^{p-k} =0\, .
\end{equation}
The $t_{k}$s depend on the $v_{i}$s and
correspond to the $p-1$ independent relevant deformations of ${\rm C}_{p}$.
They are associated with a set of critical exponents
\begin{equation}
\label{ce}
\delta^{(p)}_{k} = {1\over p-k}\, \cdotp
\end{equation}
For example, if the most relevant operator 
$t_{0}$ is turned on, then the separation of the roots is of order 
$t_{0}^{\delta^{(p)}_{0}}$.

Let us focus on the 
root $\sigma_{N}$ that goes to zero in the classical, or weak
coupling, limit $\Lambda/v_{i}\rightarrow 0$. In this limit,
$\sigma_{N}$ is given by an instanton series of the type (\ref{insser}),
\begin{equation}
\label{ins}
\sigma_{N} = \sum_{k=1}^{\infty} \tilde c_{k}\,\Lambda^{kN} ,
\end{equation}
where the coefficients $\tilde c_{k}$ can be expressed in terms of
the polynomial $P$,
\begin{equation}
\label{ck}
\tilde c_{k} = {1\over k!} \,\Bigl( {1\over \displaystyle P'}
{\d\over \d\sigma}\Bigr)^{k-1}\!\cdot 
{1\over P'}{\lower 6pt\hbox{\strut\, \vrule}}_{\, \sigma = 0} .
\end{equation}
The radius of convergence of (\ref{ins}) as a function of the distribution 
of the $v_{i}$s can be calculated at large $N$ 
with the methods of \cite{fer0b}. For our purposes, it is enough to
consider the simple case $v_{1}=\cdots=v_{N-1}=v$. In terms of the 
dimensionless ratio
\begin{equation}
\label{rdef}
r = v/\Lambda \, ,
\end{equation}
the expansion (\ref{ins}) takes the form
\begin{equation}
\label{ins2}
\sigma_{N}(r)= v\sum_{k=1}^{\infty} c_{k}\, r^{-kN}
\end{equation}
with $c_{1} = 1$, $c_{2} = 1-N$, etc\ldots
The radius of convergence of the series (\ref{ins2}) is finite because 
$\sigma_{N}(r)$ has branch cuts. The branching points occur when the 
root $\sigma_{N}$ coincides with another root of the
equation (\ref{toyex}), and thus correspond to critical points of the 
type ${\rm C}_{2}$. It is elementary to show that there are $N$
critical values $r_{*}$ of $r$ given by
\begin{equation}
\label{rcdef}
r_{*}^{N} = - {N^{N}\over (N-1)^{N-1}}
\end{equation}
and for which
\begin{equation}
\label{xcdef}
\sigma_{N}(r_{*}) = \sigma_{*} = -v/N \, .
\end{equation}
We see that $\sigma_{N}(r_{*})$ is of order $1/N$. There are 
$N$ different analytic continuations of $\sigma_{N}(r)$ for $|r|<|r_{*}|$,
corresponding to the $N$ branching points (\ref{rcdef}), and which yield 
any of the $N$ roots of (\ref{toyex}). To obtain the large $N$ expansion of 
$\sigma_{N}(r)$ for $|r|<|r_{*}|$, we write (\ref{toyex}) as
\begin{equation}
\label{Nex1}
\ln (1+x) + {1\over N} \ln {x\over 1+x} = \ln {1\over \rho_{\alpha}}\,
\cvp
\end{equation}
where $x=\sigma /v$ and $\rho_{\alpha} =e^{i\alpha}r=e^{2i\pi k/N} r$,
$0\leq k\leq N-1$. The integer $k$ 
labels the different analytic continuations. In the $1/N$ expansion, 
$\alpha = 2\pi k/N\in [0,2\pi[$ is actually a continuous variable.
At leading order, the 
term proportional to $1/N$ in (\ref{Nex1}) is negligible, and we get
$\sigma_{N}(r)/v = -1 + 1/\rho_{\alpha} + o(1)$.
Corrections are obtained by substituting $x\rightarrow 
-1 + 1/\rho_{\alpha} + \Delta$ in (\ref{Nex1}) and solving for $\Delta$. 
The first few terms are
\begin{equation}
\label{Nex2}
{\sigma_{N}(r)\over v} = -1 + {1\over \rho_{\alpha}} -
{\ln (1-\rho_{\alpha})\over 
N\rho_{\alpha}} + {1\over 2\rho_{\alpha}N^{2}} \Bigl[ \left(\strut\ln 
(1-\rho_{\alpha})\right)^{2} + {2\rho_{\alpha}\ln (1-\rho_{\alpha})\over 
1-\rho_{\alpha}} \Bigr]+ {\cal O}(1/N^{3}) .
\end{equation}

The equation (\ref{Nex2}) displays all the main features of the 
large $N$ expansion of the analytic continuations of instanton series. 
The same features are found in the full gauge theory calculations 
\cite{fer0b}. First of all, it is an asymptotic series {\it with expansion
parameter $1/N$}. This is very different from the series in $1/N^{2}$ 
that is generated by the 't~Hooft expansion in terms of Feynman 
diagrams.\footnote{Let us emphasize that the $\suN$ gauge theories we 
are considering have fields in the adjoint representation only.} It 
means that open strings must be present in any string theoretic 
description of the ${\cal N}=2$ gauge theory.
The physics underlying this fact is that the pure closed 
string background is singular at strong coupling. The singularity 
is resolved by inflating branes \cite{enh}. The open strings must be 
the strings that are attached to these branes \cite{fer0b}. 
A second general qualitative feature of (\ref{Nex2}) is that the 
contribution at each order in $1/N$ is given by a series in 
$1/\rho_{\alpha}$, by writing $\ln (1-\rho_{\alpha}) =
\ln (-\rho_{\alpha}) + \ln (1-1/\rho_{\alpha}) = \ln 
(-\rho_{\alpha}) -\sum_{j=1}^{\infty}1/(j\rho_{\alpha}^{j})$. Since
an instanton contributes $1/\rho_{\alpha}^{N}$, we 
may interpret those terms as coming from fractional instantons. The 
physical picture is then that instantons have disintegrated through
the process of analytic continuation in the strong coupling 
region.\footnote{The fractional instanton picture is only heuristic, 
because we have not found the corresponding field configurations (that 
must be singular in the original field variables), and also because at 
large $N$ the fractional topological charge is vanishingly small.}
The third qualitative feature is that the large $N$ expansion breaks
down at $\rho_{\alpha}=1$, which corresponds to the critical point
where two roots coincide.

This last feature is of course crucial 
for our purposes. The origin of the divergences is exactly as 
discussed in Section 3.1. Introducing
\begin{equation}
\label{deldef}
\epsilon = \rho_{\alpha} -1\, ,
\end{equation}
the separation of the two colliding roots goes like 
$\delta\sigma\propto \epsilon^{1/2}$ when $\epsilon\rightarrow 0$, the 
critical exponent $\delta^{(2)}_{0}=1/2$ being given by (\ref{ce}).
However, it is straightforward to see 
that in the leading large $N$ approximation
$\delta^{(2)}_{0}=1$. This erroneous result is obtained by noting that the 
formula (\ref{Nex2}) gives the large $N$ expansion of the $N-1$ roots 
$\sigma_{1},\ldots,\sigma_{N-1}$ for $\rho_{\alpha}>1$, a regime
in which $\sigma_{N} = 0$ to all orders in $1/N$ due to the 
exponential decrease of instantons. The divergences in the corrections
to the leading approximation in (\ref{Nex2}) when $\epsilon\rightarrow 
0$ signal the failure of the
$1/N$ expansion to yield the correct critical exponent, as discussed
in Section 3.1.

The next step is to check whether the divergences are specific enough 
for a double scaling limit to be defined. For this purpose,
let us consider the rescaled observable
\begin{equation}
\label{scax}
s = N \sigma_{N}/v
\end{equation}
and the double scaling limit
\begin{equation}
\label{scatoy}
N\rightarrow\infty\, ,\quad \epsilon\rightarrow 0\, ,\quad N\epsilon - \ln 
N = {\rm constant} = 1/\kappa + \ln\kappa\, .
\end{equation}
By plugging in (\ref{scatoy}) into (\ref{Nex2}), and discarding terms of 
order $\kappa^{2}$ and higher, we get
\begin{eqnarray}
\label{dstoy1}
& s = \left( \strut -1/\kappa -\ln\kappa - \ln N \right) + 
\left( \strut -\ln (-1) + \ln\kappa + \ln N - 
\kappa\ln\kappa -\kappa\ln N \right) \nonumber \\
&\hphantom{s =}
+\left( \strut -\kappa\ln (-1) + \kappa\ln N + \kappa\ln\kappa \right)
+ \cdots
\end{eqnarray}
We have grouped together in (\ref{dstoy1}) the terms that come from
a given order in $1/N$ in (\ref{Nex2}). We see that subtle 
cancellations between the different terms make the result finite to 
the order we consider,
\begin{equation}
\label{dstoy2}
s = -1/\kappa - \ln (-1) - \kappa\ln (-1) +{\cal O}(\kappa^{2})\, .
\end{equation}
It is actually very easy to see that the cancellations will work to all
orders and beyond. By using the scalings (\ref{scax}) and (\ref{scatoy}) in 
the exact equation (\ref{Nex1}), we indeed obtain a non-perturbative 
equation determining $s$,
\begin{equation}
\label{nps}
se^{s} = {1\over\kappa}\, e^{-1/\kappa}\, .
\end{equation}
This is an {\it exact} result for the double scaled theory, from 
which we can in particular derive the full asymptotic series in 
$\kappa$ and thus recover (\ref{dstoy2}). 
The analogous string theoretic results that we obtain by using a gauge 
theory instead of the simple toy model (\ref{toyex}) are described 
below.
\subsection{Exact results in 4D string theory}
For concreteness, let us write down explicitly some of the 
exact results obtained in \cite{fer1}. The critical points we consider are 
Argyres-Douglas critical points ${\rm AD}_{p}$ which have $p-1$ relevant 
deformations $t_{k}$, $0\leq k\leq p-2$. The space-time dimensions of the 
$t_{k}$s can be deduced from the exact solution for the parent ${\cal N}=2$ 
supersymmetric gauge theory, and read
\begin{equation}
\label{stdim}
[t_{k}]_{\rm space-time} = {2(p-k)\over p+2}\,\cdotp
\end{equation}
The double scaling limit, similar to (\ref{dsca2}), is
\begin{equation}
\label{sca1a}
N\rightarrow\infty\, ,\quad t_{k}\rightarrow 0\, ,\quad
\tau_{k} = N^{1-k/n}\,t_{k} = {\rm constant}\, .
\end{equation}
By identifying the most relevant operator $\tau_{0}$ with the world sheet 
cosmological constant, and
by using the reasoning in Section 2.4, equation (\ref{ccren}) and below,
we deduce that the world sheet cut-off $1/\ell$ scales as
\begin{equation}
\label{wscutoffsca}
\ell^{2} \sim 1/N
\end{equation}
and that the world sheet dimensions are
\begin{equation}
\label{wsdim}
[\tau_{k}]_{\rm world\ sheet} = {2(p-k)\over p}\,\cdotp
\end{equation}
The string theoretic central charge $z$ of the supersymmetry algebra 
as a function of the world sheet couplings $\tau_{k}$ reads
\begin{equation}
\label{ccstring}
z(\gamma) = {1\over 4\tau_{0}^{1/p}}\, \oint_{\gamma}
{\displaystyle u T'(u)\over\displaystyle\sqrt{1-e^{-T(u)}}}\, \d u\, ,
\end{equation}
with
\begin{equation}
\label{Tdef}
T(u) = \sum_{k=0}^{p-2}\tau_{k}u^{k} + u^{p}\, .
\end{equation}
The contour $\gamma$ encircles any two roots of the polynomial $T$, and 
corresponds to a choice of electric and magnetic charges. The formula 
(\ref{ccstring}) includes an infinite series of string perturbative 
corrections as well as all the non-perturbative contributions.
It can be viewed as the ``realistic'' generalization of the toy model
equation (\ref{nps}). For 
example, in the special case where only the most 
relevant operator is turned on, it is natural to introduce the 
coupling $\kappa = 1/\tau_{0}$, to rescale 
$u\rightarrow \kappa^{-1/p}u$, and to consider the contours $\gamma_{jk}$ 
encircling the roots
$u_{j}=\exp (i\pi (1 + 2j)/p)$ and $u_{k}$. A straightforward 
calculation then yields
\begin{equation}
\label{Ai1}
z(\gamma_{jk}) = {p\over 4\kappa}\,\oint_{\alpha_{jk}}
{\displaystyle u^{p}\, 
\d u\over\displaystyle\sqrt{1-e^{-(1+u^{p})/\kappa}}} =
e^{i\pi (j+k+1)/p}\sin\bigl(\pi(j-k)/p\bigr)\, {\cal I}_{p}(\kappa)
\end{equation}
where
\begin{equation}
\label{Indef}
{\cal I}_{p}(\kappa) = {p\over (p+1)\kappa} + \int_{0}^{1/\kappa}
\Bigl( {\displaystyle1\over\displaystyle\sqrt{1-e^{-x}}}-1 \Bigr) 
(1-\kappa x)^{1/p}\d x.
\end{equation}
The asymptotic expansion of ${\cal I}_{p}(\kappa)$ can be obtained by 
noting that when $x\sim 1/\kappa$, $1/\sqrt{1-\exp(-x)} -1$ is 
exponentially small. We thus have
\begin{equation}
\label{Inas}
{\cal I}_{p}(\kappa) = {p\over (p+1)\kappa} + \sum_{k=0}^{K}
{\Gamma(k-1/p)\over\Gamma(-1/p)\Gamma(k+1)} I_{k}\, \kappa^{k} + {\cal 
O}(\kappa^{K+1})
\end{equation}
with
\begin{equation}
\label{Idef2}
I_{k} = \int_{0}^{\infty}\Bigl( {\displaystyle 1\over\displaystyle\sqrt{1
-e^{-x}}} - 1 \Bigr) x^{k}\, \d x = {(-1)^{k+1}\over k+1}\,
\int_{0}^{1}{\left( \ln(1-t) \right)^{k+1}\over 2t^{3/2}}\, \d t.
\end{equation}
The first integrals $I_{k}$ can be calculated by expanding the logarithm 
in powers of $t$,
\begin{equation}
\label{expik}
I_{0}=2\ln2\, ,\quad I_{1} = {\pi^{2}\over 6} - 2 (\ln 2)^{2}\, ,\quad
I_{2} = {8 (\ln 2)^{2}\over 3} - {2 \pi^{2}\ln 2\over 3} + 4 \zeta(3)\, ,
\end{equation}
which gives the first string loops corrections.

An important point is that the central charge $Z$, and thus the BPS masses,
of the parent gauge theory scales as \cite{fer1}
\begin{equation}
\label{Zscagt}
Z \sim N^{-1/p} z\sim \ell^{2/p}\, ,
\end{equation}
and thus the continuum UV limit on the world sheet $\ell\rightarrow 0$ 
corresponds to a low energy limit of the parent gauge theory $Z\rightarrow 
0$. This shows explicitly that in the double scaling limit we are left with 
the low energy degrees of freedom only, as was already deduced from 
general arguments at the end of Section 3.1.
\subsection{Further insights}
\subsubsection{Full proofs}
The consideration of the protected observables on the gauge theory 
side, that correspond to the central charge or equivalently to the 
low energy effective action, is not enough to 
give a full proof of the existence of
double scaling limits. One should study in principle {\it all}
the observables, including those with a non-trivial perturbative 
expansion, or equivalently the full path integral.
Moreover, the heuristic picture for the appearance of
a continuum string theory in the limit relies on the observation
that very large Feynman graphs dominate near the critical points, as 
reviewed in Section 2.4. This can in principle
be checked on generic amplitudes but obviously
not on the BPS observables for which perturbation theory is 
trivial. One should also give a proof that the continuum 
limit does correspond to a genuine continuum string theory, a fact 
that is extremely difficult to check directly even in the $D\leq 1$ cases.

Unfortunately, generic amplitudes cannot be calculated in 
Yang-Mills theory. For that reason, it is interesting to consider 
simplified models that can be exactly solved. A particularly 
interesting one was studied in \cite{fer2}. The model is a two 
dimensional non-linear $\sigma$ model
which is a very close relative to ${\cal N}=2$ super Yang-Mills in 
four dimensions. It has an exactly calculable central charge 
with the same non-renormalization theorems as in four dimensions and 
the same BPS mass formula. The analysis sketched in Sections 3.1--3.4 
can thus be reproduced, with qualitatively the same
results (appearance of `fractional instantons,' breakdown of the 
large $N$ expansion at critical points, possibility to define double 
scaling limits for which exactly known BPS amplitudes
have a finite limit). Moreover, and this is the main point, the
two-dimensional model is exactly solvable in the large $N$ limit: the 
large $N$ Feynman graphs can be explicitly summed up.
It is then possible to give rigorous 
proofs of the existence of the double scaling limits, and to fully 
characterize the double scaled theories. In particular, we do obtain
the expected continuum limits.
Those results are extremely encouraging and strongly suggest that the 
scaling limits are consistent in the case of the gauge theories as 
well. We invite the reader to consult \cite{fer2} for more details.
\subsubsection{Non-perturbative non-Borel summable partition functions}
As we have already emphasized, a very important point of principle is 
that the four dimensional double scaling limits are non-perturbative. 
This is to be contrasted with the classic $D\leq 1$ cases, 
for which the non-Borel summable observables 
are not defined beyond perturbation theory. We have discussed this aspect
in some details in \cite{fer3}, where interesting mechanisms that allow for a 
non-perturbative definition of non-Borel summable partition functions 
are explicitly worked out in a class of simple theories 
akin to the model studied in \cite{fer2}. 
\section{Open problems}
The results obtained in \cite{fer1} show that a generalization
of the discretized approach to four dimensional non-critical strings
is possible. However, we have only scratched the surface
of the subject, and many points would deserve further
investigations. For example, the critical points used in \cite{fer1}
only correspond to one family amongst many others that can be
found on the moduli space of supersymmetric gauge theories. A
classification of these critical points exists \cite{AD}, with an ADE 
pattern, and the
work in \cite{fer1} could certainly be generalized to the most general
cases. Another interesting line of research is the study of
the renormalization group flow
equations between the different critical points that can be
derived by using the explicit formulas of \cite{fer1}.
In the classic low dimensional cases, a very
elegant mathematical structure emerges (generalized KdV hierarchy)
when one studies the flows (see for example Sections 4 and 8 of
\cite{revZJ}). It would be extremely interesting to discover a similar
structure in the four dimensional theories. Yet another intriguing 
point is that the approach of
\cite{fer1} suggests a relationship between the analytic continuation
of the sum over gauge theory instantons in the double scaling limit
and sums over Riemann surfaces with boundaries which define the
perturbative series of the resulting string theories. It would be
worth studying if this correspondence can be proven directly by
looking at the moduli space of instantons at large $N$ and large
instanton number $k$. Finally, it is highly desirable to construct
directly in the continuum the string theories obtained in the double
scaling limits. As in the classic $D\leq 1$ case, this would provide 
an important consistency check of the discretized approach. Moreover, 
we believe that the ADE `exactly solvable' four dimensional 
non-critical strings could play a r\^ole similar to the ADE minimal 
two dimensional CFTs, and shed considerable light into the structure of 
non-perturbative string theories.
\section*{Acknowledgements}
I would like to thank the Les Houches summer school for providing a 
very nice learning and thinking environment. I am grateful to Rafael 
Hern\'andez, Igor Klebanov, Ivan Kostov
and particularly Alexandre Polyakov for 
comments on the work presented in these lectures. The research was 
supported by a Robert H.~Dicke fellowship from Princeton University, 
and more recently by the Swiss National Science Foundation and the 
University of Neuch\^atel.
\end{document}